\documentclass[preprintnumbers,amsmath,amssymbm,prd]{revtex4}
\usepackage{epsfig}
\usepackage{graphicx}
\usepackage{amssymb}

\begin{document}
\title{The quantum emission spectra of rapidly-rotating Kerr black holes: discrete or continuous?}
\author{Shahar Hod}
\address{The Ruppin Academic Center, Emeq Hefer 40250, Israel}
\address{ }
\address{The Hadassah Institute, Jerusalem 91010, Israel}
\date{\today}

\begin{abstract}
\ \ \ Bekenstein and Mukhanov (BM) have suggested that, in a quantum
theory of gravity, black holes may have discrete emission spectra.
Using the time-energy uncertainty principle they have also shown
that, for a (non-rotating) Schwarzschild black hole, the natural
broadening $\delta\omega$ of the black-hole emission lines is
expected to be small on the scale set by the characteristic
frequency spacing $\Delta\omega$ of the spectral lines:
$\zeta^{\text{Sch}}\equiv\delta\omega/\Delta\omega\ll1$. BM have
therefore concluded that the expected discrete emission lines of the
quantized Schwarzschild black hole are unlikely to overlap. In this
paper we calculate the characteristic dimensionless ratio
$\zeta(\bar a)\equiv\delta\omega/\Delta\omega$
%between the natural broadening $\delta\omega$ and the frequency spacing $\Delta\omega$
for the predicted BM emission spectra of rapidly-{\it rotating} Kerr
black holes (here $\bar a\equiv J/M^2$ is the dimensionless angular
momentum of the black hole). It is shown that $\zeta(\bar a)$ is an
{\it increasing} function of the black-hole angular momentum.
%$\bar a$ with the asymptotic property $\zeta(\bar a\to 1)\to\infty$.
In particular, we find that the quantum emission lines of Kerr black
holes in the regime $\bar a\gtrsim 0.9$ are characterized by the
dimensionless ratio $\zeta(\bar a)\gtrsim1$ and are therefore
effectively blended together. Our results thus suggest that, even if
the underlying mass (energy) spectrum of these rapidly-rotating Kerr
black holes is fundamentally {\it discrete} as suggested by
Bekenstein and Mukhanov, the natural broadening phenomenon
(associated with the time-energy uncertainty principle) is expected
to smear the black-hole radiation spectrum into a {\it continuum}.
\end{abstract}
\bigskip
\maketitle

%]

\section{Introduction}

Analyzing the quantum properties of fundamental fields in classical
black-hole spacetimes, Hawking \cite{Haw1} has revealed that black
holes are actually not completely black. In particular, according to
Hawking's result, semi-classical black holes are characterized by
{\it continuous} evaporation spectra in which the emitted field
quanta have the familiar black-body statistical distribution with a
well defined temperature \cite{Haw1,Notewor}. This intriguing
theoretical prediction is certainly one of the most important
outcomes of the interplay between quantum field theory and classical
general relativity.

It is important to realize, however, that Hawking's seminal analysis
\cite{Haw1} has a {\it semi-classical} nature: while the fundamental
fields are properly analyzed at the {\it quantum} level, the curved
black-hole spacetime is treated as a {\it classical} background.
Although we do not have a self-consistent theory of quantum gravity,
it is natural to expect that some modifications to the
(semi-classically predicted) Hawking emission spectrum \cite{Haw1}
may arise if the black-hole spacetime itself would properly be
treated as a dynamical {\it quantum} entity \cite{Beken1}.

An interesting heuristic quantization of the black-hole area (mass)
spectrum was proposed long ago by Bekenstein \cite{Beken1}. In his
influential work, Bekenstein has pointed out that the black-hole
surface area behaves as a classical adiabatic invariant
\cite{Beken1}. Since classical adiabatic invariants are usually
related to physical quantities with discrete quantum spectra
\cite{Ehr}, Bekenstein has suggested that the black-hole surface
area has a discrete (quantum) spectrum of the form
\cite{Beken1,Noteunit}
\begin{equation}\label{Eq1}
A_n=4\gamma\hbar\cdot n\ \ \ ;\ \ \ n=1,2,3,\ldots\ \  .
\end{equation}
Here $\gamma$ is a dimensionless constant of order unity. Three
possible values of $\gamma$ are often used in the literature:
$\gamma=2\pi$ \cite{Beken1}, $\gamma=\ln2$ \cite{BekMuk,Hodch}, and
$\gamma=\ln3$ \cite{Hodln}. It is worth mentioning that, using
different quantization schemes, several authors (see
\cite{BekMuk,Hodch,Hodln,Muk,Kog,Maz,Mag,Lou,Pel,Louk,Bar,Kas,MakRep,HodD,HodG,Boj,Alu,Gar1,Gar2,Magim,Kund,Kes}
and references therein) have re-derived the uniformly spaced
black-hole area spectrum (\ref{Eq1}).

The discrete black-hole area spectrum (\ref{Eq1}) also implies a
discrete mass (energy) spectrum $\{M_n\}$ for quantum black holes.
Thus, as pointed out in \cite{Beken1}, a quantized black hole is
expected to be characterized by a {\it discrete} line emission. In
particular, Bekenstein and Mukhanov (BM) \cite{BekMuk} have
advocated the idea that, within the framework of a quantum theory of
gravity, the radiation emitted by a quantized Schwarzschild black
hole of mass $M_n$ should be at integer multiples of the fundamental
frequency \cite{Beken1,BekMuk}
\begin{equation}\label{Eq2}
\omega_0\equiv (M_n-M_{n-1})/\hbar={{\gamma
T^{\text{Sch}}_{\text{BH}}}\over{\hbar}}\ ,
\end{equation}
where
\begin{equation}\label{Eq3}
T^{\text{Sch}}_{\text{BH}}={{\hbar}\over{8\pi M}}
\end{equation}
is the Bekenstein-Hawking temperature \cite{Haw1,Beken1} of the
Schwarzschild black hole.

According to the quantization scheme presented in
\cite{Beken1,BekMuk}, the decay of a quantized Schwarzschild black
hole of mass $M_n$ into the $M_{n-k}$ mass level \cite{Notenk} is
accompanied by the emission of a field quantum of frequency
$\omega_k=(M_n-M_{n-k})/\hbar=k\cdot\omega_0$ [see Eq. (\ref{Eq1})].
This implies that the discrete line emission
$\{\omega_0,2\omega_0,3\omega_0,...\}$ suggested by BM
\cite{Beken1,BekMuk} for a quantized Schwarzschild black hole is
characterized by the constant frequency spacing
\begin{equation}\label{Eq4}
\Delta\omega^{\text{Sch}}=\omega_0\
\end{equation}
between adjacent emission lines.

\section{The natural broadening of the Schwarzschild spectral lines}

The interesting question of the {\it natural broadening} of the
Schwarzschild black-hole emission lines has been addressed by BM
\cite{BekMuk} (see also \cite{Mak,Hod1}). Arguing from the
time-energy uncertainty principle, it was suggested in \cite{BekMuk}
to estimate the natural broadening $\delta\omega$ of the
Schwarzschild black-hole spectral lines from the relation
$\delta\omega=1/\tau$ \cite{Ehr}, where $\tau$ is the characteristic
lifetime (as measured by asymptotic observers) of the $n$th
black-hole mass (energy) level \cite{Notema}.

As suggested by BM \cite{BekMuk}, the characteristic lifetime,
$\tau$, of the Schwarzschild $n$th mass level can be estimated from
the relation $\tau=(dN/dt)^{-1}$, where $dN/dt$ is the
semi-classical emission rate of the black hole
\cite{Haw1,Page,Notecor,Jack,Hor}. For the emission of (massless)
gravitons and photons from a Schwarzschild black hole one finds the
emission rate $(dN/dt)^{\text{Sch}}\simeq 1.6\times 10^{-4}M^{-1}_n$
\cite{Page}, which implies $\tau^{\text{Sch}}\simeq6\times 10^3M_n$
\cite{Noterec} for the survival time (lifetime) of the $n$th
Schwarzschild mass level. Using the time-energy uncertainty
relation, $\delta\omega=1/\tau$ \cite{Ehr}, one finds
\begin{equation}\label{Eq5}
\delta\omega^{\text{Sch}}\simeq1.6\times 10^{-4}M^{-1}
\end{equation}
for the characteristic natural broadening of the Schwarzschild
spectral lines.

Taking cognizance of Eqs. (\ref{Eq2})-(\ref{Eq5}), one concludes
that, for a Schwarzschild black hole \cite{BekMuk}, the frequency
spacing $\Delta\omega^{\text{Sch}}$ and the natural width
$\delta\omega^{\text{Sch}}$ of the spectral lines are well separated
in magnitude. In particular, one finds [see Eqs.
(\ref{Eq2})-(\ref{Eq5})]
\begin{equation}\label{Eq6}
\zeta^{\text{Sch}}\equiv{{{\delta\omega}^{\text{Sch}}}\over{{\Delta\omega}^{\text{Sch}}}}\simeq
4\times 10^{-3}\ll 1\ .
\end{equation}
The strong inequality (\ref{Eq6}) implies that the discrete
Schwarzschild black-hole emission lines
$\{\omega_0,2\omega_0,3\omega_0,...\}$ suggested by BM
\cite{Beken1,BekMuk} are unlikely to overlap.

\section{The spectral emission lines of rotating Kerr black holes and their natural broadening}

In this paper we shall generalize the analyzes of
\cite{BekMuk,Mak,Hod1} to the regime of {\it rotating} Kerr black
holes. In particular, our main goal is to study the dependence of
the fundamental dimensionless ratio $\zeta(\bar a)\equiv
\delta\omega/\Delta\omega$ on the black-hole rotation parameter
$\bar a\equiv J/M^2$ \cite{Notejj}.

\subsection{The quantized emission spectrum of the Kerr black hole}

The Bekenstein-Hawking temperature $T_{\text{BH}}$ and the angular
velocity $\Omega_{\text{H}}$ of a rotating Kerr black hole are
respectively given by the relations \cite{Haw1,Beken1,Noteunit}
\begin{equation}\label{Eq7}
T_{\text{BH}}={{\hbar(r_+-r_-)}\over{4\pi(r^2_++a^2)}}\ \ \ \
\text{and}\ \ \ \ \Omega_{\text{H}}={{a}\over{2Mr_+}}\  ,
\end{equation}
where
\begin{equation}\label{Eq8}
r_{\pm}=M\pm (M^2-a^2)^{1/2}
\end{equation}
are the black-hole horizon radii.

Taking cognizance of the discrete black-hole area spectrum
(\ref{Eq1}) suggested by Bekenstein \cite{Beken1}, and using the
first law of black-hole mechanics \cite{Notequ}
\begin{equation}\label{Eq9}
\Delta M={{1}\over {4\hbar}}T_{\text{BH}}\Delta
A+\Omega_{\text{H}}\Delta J\ ,
\end{equation}
one finds that a quantized rotating Kerr black hole is expected to
be characterized by the discrete emission frequencies \cite{Notepos}
\begin{equation}\label{Eq10}
\omega_{k,m}={{\gamma T_{\text{BH}}}\over{\hbar}}\cdot
k+m\Omega_{\text{H}}\ \ \ ;\ \ \ k\in\mathbb{Z}\ \  .
\end{equation}
Here $m$ is the azimuthal harmonic index of the emitted field
quanta. Note that the discrete black-hole radiation spectrum
(\ref{Eq10}) is characterized by the
%constant
frequency spacing \cite{Noterr}
\begin{equation}\label{Eq11}
\Delta\omega=\omega_{k+1,m}-\omega_{k,m}={{\gamma
T_{\text{BH}}}\over{\hbar}}\
\end{equation}
%between adjacent emission lines.

\subsection{The natural broadening of the Kerr spectral lines}

Following the Bekenstein-Mukhanov analysis of the Schwarzschild
black-hole emission spectrum presented in \cite{BekMuk}, we shall
now use the time-energy uncertainty principle \cite{Ehr} in order to
estimate the natural broadening $\delta\omega=1/\tau$ of the Kerr
black-hole emission lines. In particular, following \cite{BekMuk} we
shall assume that the characteristic lifetime $\tau$ of the Kerr
$n$th mass level (that is, the average time gap between quantum
leaps) can be estimated as the reciprocal of the semi-classical
\cite{Haw1,Page,Notecor,Jack,Hor} black-hole emission rate [see Eq.
(\ref{Eq15}) below] \cite{Notehb}.

The semi-classical emission rate of a rotating Kerr black hole (that
is, the number of quanta emitted from the black hole per unit of
time) is given by the Hawking relation \cite{Haw1,Page,Notepos2}
\begin{equation}\label{Eq12}
{\cal \dot
N}\equiv{{dN}\over{dt}}={{1}\over{2\pi}}\sum_{s,l,m}\int_0^{\infty}
d\omega{{\Gamma}\over{e^{\hbar(\omega-m\Omega_{\text{H}})/T_{\text{BH}}}-1}}\
.
\end{equation}
%where $x\equiv\hbar(\omega-m\Omega_{\text{H}})/T_{\text{BH}}$.
Here $s, m$ and $l\geq \max(s,|m|)$ are respectively the spin
parameter and the harmonic indices (azimuthal and spheroidal) of the
emitted field quanta. The energy-dependent grey-body factors
(absorption probabilities) $\Gamma=\Gamma_{slm}(\omega;\bar a)$
\cite{Page} in (\ref{Eq12}) quantify the interaction of the emitted
field quanta with the effective curvature potential that surrounds
the emitting black hole.

The dependence of the semi-classical emission rate ${\cal \dot
N}(\bar a)$ [see Eq. (\ref{Eq12})] on the black-hole rotation
parameter $\bar a$ can be computed along the lines of the numerical
procedure described in \cite{Page}. In particular, one finds that,
in the regime of rapidly-rotating Kerr black holes, the
semi-classical radiation spectrum of a massless spin-$s$ field is
greatly dominated by the \cite{Page,Notegsr}
\begin{equation}\label{Eq13}
l=m=s\
\end{equation}
angular mode. In addition, the characteristic thermal (exponential)
factor that appears in the expression (\ref{Eq12}) for the
black-hole emission rate implies that, in the regime of
rapidly-rotating (near-extremal, $T_{\text{BH}}\to 0$) black holes,
the emission of high energy quanta with $\omega>m\Omega_{\text{H}}$
is exponentially suppressed. Thus, the semi-classical emission
spectra of rapidly-rotating Kerr black holes are dominated by field
quanta in the energy interval \cite{Noteexd}
\begin{equation}\label{Eq14}
0\leq\omega\lesssim m\Omega_{\text{H}}+O(T_{\text{BH}}/\hbar)\  .
\end{equation}

As discussed above, following \cite{BekMuk} we shall assume that the
lifetime $\tau$ of the meta-stable black-hole state (that is, the
average time gap between the emissions of successive black-hole
quanta) is given by the reciprocal of the black-hole semi-classical
emission rate \cite{Notecor}. Namely,
\begin{equation}\label{Eq15}
\tau={{1}\over{{\cal \dot N}}}\  ,
\end{equation}
where ${\cal \dot N}$ is given by Eq. (\ref{Eq12}). Using the
time-energy uncertainty principle \cite{Ehr}, one finds (see also
\cite{BekMuk})
\begin{equation}\label{Eq16}
\delta\omega={{1}\over{\tau}}={{\cal \dot N}}\
\end{equation}
for the natural broadening of the black-hole emission lines.

In Table \ref{Table1} we display the dimensionless ratio $\zeta(\bar
a)\equiv\delta\omega/\Delta\omega$ \cite{Notegam} for the emission
of massless gravitons ($s=2$) and photons ($s=1$)
\cite{Notegsr,Notesup} by rapidly-rotating Kerr black holes [Here
the natural broadening $\delta\omega$ of the black-hole spectral
lines is given by Eqs. (\ref{Eq12}) and (\ref{Eq16}), and the
characteristic frequency spacing $\Delta\omega$ between adjacent
emission lines \cite{Noterr} is given by (\ref{Eq11})]. One finds
that $\zeta(\bar a)$ is an {\it increasing} function of the
dimensionless black-hole rotation parameter $\bar a$. In particular,
we find that rapidly-rotating Kerr black holes in the regime $\bar
a\gtrsim 0.9$ are characterized by the relation
\cite{Noteex,Noteasy}
\begin{equation}\label{Eq17}
\delta\omega\gtrsim\Delta\omega\  .
\end{equation}
The inequality (\ref{Eq17}) implies that the emission lines of
rapidly-rotating Kerr black holes are effectively blended together.

\begin{table}[htbp]
\centering
\begin{tabular}{|c|c|c|c|c|c|}
\hline $\ \ \ \bar a\equiv J/M^2\ \ \ $ & \ \ \ 0.80 \ \ \ & \ \ \
0.90 \ \ \ & \ \ \ 0.96 \ \ \ & \ \ \ 0.99
\ \ \ & \ \ \ 0.999\ \ \ \ \\
\hline \ \ $\delta\omega/\Delta\omega$
\ \ &\ 0.12\ \ &\ 0.39\ \ &\ 1.34\ \ &\ 5.39\ \ &\ \ 27.98\ \ \ \\
\hline
\end{tabular}
\caption{The characteristic dimensionless ratio $\zeta(\bar
a)\equiv\delta\omega/\Delta\omega$ of rapidly-rotating Kerr black
holes. Here $\delta\omega$ is the natural broadening of the
black-hole emission lines [see Eqs. (\ref{Eq12}) and (\ref{Eq16})]
and $\Delta\omega$ \cite{Noterr,Notegam} is the characteristic
frequency spacing between adjacent emission lines [see Eq.
(\ref{Eq11})]. One finds that $\zeta(\bar a)$ is an {\it increasing}
function of the black-hole angular momentum. In particular, we find
that the dimensionless ratio $\delta\omega/\Delta\omega$ becomes of
order unity at $\bar a\simeq0.9$ \cite{Noteex,Noteasy}.}
\label{Table1}
\end{table}

\section{Summary}

Starting with the seminal work of Bekenstein \cite{Beken1}, many
authors (see
\cite{BekMuk,Hodch,Hodln,Muk,Kog,Maz,Mag,Lou,Pel,Louk,Bar,Kas,MakRep,HodD,HodG,Boj,Alu,Gar1,Gar2,Magim,Kund,Kes}
and references therein) have predicted the existence of a uniformly
spaced area spectrum [see Eq. (\ref{Eq1})] for quantized black
holes. This intriguing prediction suggests that, in a quantum theory
of gravity \cite{Noteqg}, black holes may have {\it discrete}
emission spectra.

Using the time-energy uncertainty principle \cite{Ehr}, Bekenstein
and Mukhanov \cite{BekMuk} have shown that, for ({\it non}-rotating)
Schwarzschild black holes, the natural broadening
$\delta\omega^{\text{Sch}}$ of the spectral lines is small on the
scale set by the characteristic frequency spacing
$\Delta\omega^{\text{Sch}}$ of the lines:
$\delta\omega^{\text{Sch}}/\Delta\omega^{\text{Sch}}\ll1$. It was
therefore concluded in \cite{BekMuk} that the discrete emission
lines which are expected to characterize a quantized Schwarzschild
black hole are unlikely to overlap. In other words, the
Schwarzschild line spectrum is expected to be sharp.

Motivated by this important conclusion, in this paper we have
analyzed the natural broadening of the emission lines which,
according to \cite{Beken1}, are expected to characterize quantized
{\it rotating} Kerr black holes. In particular, we have studied the
dependence of the dimensionless ratio $\zeta(\bar
a)\equiv\delta\omega/\Delta\omega$ on the black-hole rotation
parameter $\bar a\equiv J/M^2$. It was shown that $\zeta(\bar a)$ is
an {\it increasing} function of the black-hole angular momentum. In
particular, one finds that the quantum emission lines of
rapidly-rotating Kerr black holes in the regime $\bar a\gtrsim 0.9$
are characterized by the dimensionless ratio $\zeta(\bar a)\gtrsim1$
\cite{Noteex,Noteasy}. These emission lines are therefore
effectively blended together.

Our results thus suggest that, even if the underlying mass (energy)
spectrum of these rapidly-rotating Kerr black holes is fundamentally
{\it discrete} as suggested by Bekenstein and Mukhanov, the natural
broadening phenomenon (associated with the time-energy uncertainty
principle \cite{Ehr}) is expected to smear the black-hole emission
spectrum into a {\it continuum}.

\bigskip
\noindent
{\bf ACKNOWLEDGMENTS}
\bigskip

This research is supported by the Carmel Science Foundation. I thank
Yael Oren, Arbel M. Ongo, Ayelet B. Lata, and Alona B. Tea for
stimulating discussions.

%\newpage


\begin{thebibliography}{99}
\bibitem{Notewor} It is worth noting that the thermal black-hole radiation
spectrum is distorted by the scattering potential (the effective
curvature potential) which surrounds the emitting black hole.
\bibitem{Noteunit} We use natural units in which $G=c=k_{\text{B}}=1$.
\bibitem{Notema} That is, $\tau$ is the mean time between quantum transitions of the black
hole from level to level.
\bibitem{Noterec} Here we have used the relation $\tau\equiv (dN/dt)^{-1}$
for the mean time between quantum leaps.
\bibitem{Notejj} Here $J$ is the angular momentum of the Kerr black
hole.
\bibitem{Notequ} Here we have used the area-entropy relation
$S_{\text{BH}}=A/4\hbar$ for black holes.
\bibitem{Notepos} Note that only positive frequency $\omega\geq0$ modes
(that is, field quanta with positive energy as measured by
asymptotic observers) can be emitted by the black hole (that is, can
reach infinity). Thus, the allowed values of the discrete parameters
$\{k,m\}$ are restricted by the requirement $\omega_{k,m}={{\gamma
T_{\text{BH}}}/{\hbar}}\cdot k+m\Omega_{\text{H}}\geq0$.
\bibitem{Noterr} We are interested in rapidly-rotating Kerr
black holes with $a=O(M)$. These black holes are characterized by
the relation [see Eq. (\ref{Eq7})]
$T_{\text{BH}}/\hbar\ll\Omega_{\text{H}}$. Note that this strong
inequality implies [see Eq. (\ref{Eq10})]
$\omega_{k+1,m}-\omega_{k,m}\ll \omega_{k,m+1}-\omega_{k,m}$ (and,
in particular, $\omega_{k,0}\ll \omega_{0,1}$ in the regime $k\cdot
T_{\text{BH}}/\hbar\ll\Omega_{\text{H}}$) for these rapidly-rotating
Kerr black holes.
\bibitem{Notehb} It is worth noting that there is one important
difference between our approach for the calculation of the Kerr
black-hole emission rate and the closely related approach used in
\cite{BekMuk} for the calculation of the Schwarzschild black-hole
emission rate. In \cite{BekMuk} the authors modeled the radiating
black hole as a perfect black-body emitter and used the
Stefan-Boltzmann law in order to approximate the true radiation
power of the Schwarzschild black hole. Below we shall consider the
actual emission rates of Kerr black holes by properly taking into
account the frequency-dependent grey-body factors [see Eq.
(\ref{Eq12}) below] which quantify the partial scattering of the
emitted field quanta by the effective curvature potential of the
black-hole spacetime.
\bibitem{Notepos2} As emphasized earlier \cite{Notepos}, only positive frequency (energy)
modes (as measured by asymptotic observers) can be emitted by the
black hole \cite{Haw1,Page}. Thus, the integral in (\ref{Eq12}) is
restricted to the regime $\omega\geq0$ \cite{Haw1,Page}.
\bibitem{Notegsr} In particular, one finds \cite{Page} that, for rapidly-rotating black holes,
the semi-classical emission spectra are dominated by the
gravitational ($s=2$) field quanta with $l=m=s=2$.
\bibitem{Noteexd} Note that the exponential factor that appears in the denominator of (\ref{Eq12})
implies that, for a given value of $\omega$, the emission of modes
with $m\Omega_{\text{H}}<0$ is exponentially suppressed as compared
to the emission of modes with $m\Omega_{\text{H}}>0$. In particular,
for rapidly-rotating black holes with $\hbar
|m\Omega_{\text{H}}|/T_{\text{BH}}\gg1$, the suppression in the
emission of modes with negative values of $m\Omega_{\text{H}}$ is
described by the exponential factor
$e^{-\hbar|m\Omega_{\text{H}}|/T_{\text{BH}}}\ll1$.
\bibitem{Notegam} Here we take $\gamma=1$ in the expression (\ref{Eq11}) for $\Delta\omega$. This
order of magnitude is consistent with the most commonly used values
of the dimensionless parameter $\gamma$, see
\cite{Beken1,BekMuk,Hodch,Hodln}.
\bibitem{Noteex} Note that the exact value of $\bar a$ at which
$\delta\omega/\Delta\omega=1$ depends on the chosen value
\cite{Beken1,BekMuk,Hodch,Hodln} of the dimensionless parameter
$\gamma$.
\bibitem{Noteasy} In fact, from (\ref{Eq11}) one finds the asymptotic relation
$\zeta(\bar a\to 1)\to\infty$. Note that this asymptotic behavior is
{\it independent} of the value of the dimensionless parameter
$\gamma$.
\bibitem{Noteqg} It is worth emphasizing that by ``a quantum theory of
gravity" we mean a theory in which the black hole itself is properly
quantized. This should be contrasted with the semi-classical
analysis of Hawking \cite{Haw1}, in which the matter fields are
properly analyzed at the {\it quantum} level, but the black-hole
spacetime itself is treated as a {\it classical} background.
\end{thebibliography}
\end{document}